\documentclass[twocolumn,showpacs,superscriptaddress,floatfix,aps,prl]{revtex4-1}

\usepackage{amsmath}
\usepackage{graphicx}
\usepackage{times}
\usepackage{color}

\begin{document}

\draft

\author{U. Bortolozzo}

\author{S. Residori}
\affiliation{INLN, Universit\'e de Nice Sophia-Antipolis, CNRS, 1361 route
des Lucioles 06560 Valbonne, France}

\author{P. Sebbah}
\email[Contact: ]{sebbah@unice.fr}
\affiliation{LPMC, Universit\'e de Nice Sophia-Antipolis, CNRS, Parc Valrose
06108 Nice Cedex 2, France}

\date{\today}
\title{Experimental observation of speckle instability in Kerr random media}

\begin{abstract}
In a disordered medium with Kerr-type nonlinearity, the transmitted speckle pattern was predicted to become unstable, as a result of the positive feedback between the intensity fluctuations and the nonlinear dependence of the local refractive index. We present the first experimental evidence of speckle instability of light transversally scattered in a liquid crystal cell. A two-dimensional controlled disorder is imprinted in the medium through a suitable illumination of a photoconductive wall of the cell, whereas the nonlinear response is obtained through optical reorientation of the liquid crystal molecules. Speckle patterns with oscillating intensity are observed above a critical threshold, which is found to depend
on the scattering mean free path, as predicted by theory.
\end{abstract}


\maketitle
In a disordered medium, the transmitted intensity measured at a point results from the interference of light incoming from all possible scattering trajectories inside the medium. These constructive and destructive interferences give rise to intensity fluctuations \cite{Goodman}. The resulting speckle pattern observed in transmission is a fingerprint of the medium traversed by the wave and is highly sensitive for instance to any displacement of scatterers within the medium. This property has been widely used in applications such as speckle
imaging \cite{Komatsu77} or diffuse wave spectroscopy \cite{Maret87,Pine88}. Short and long range speckle correlations, which can also results from frequency or polarization shift of the incident beam, have raised considerable interest in the last two decades \cite{Genack09}. But changes in speckle pattern can also be achieved by intensity variation of the input beam when considering a nonlinear scattering medium with a Kerr-type nonlinearity. Local intensity fluctuations induce local refractive-index changes, which in turn modify the optical
trajectories, resulting in a redistribution of the intensity fluctuations within the sample and in a new transmitted speckle pattern. Under certain conditions, this feedback mechanism was predicted to become unstable leading to the spontaneous oscillations of the speckle pattern \cite{Spivak00,SkipetrovPRL00}. It was pointed out in \cite{SkipetrovPRL00} that speckle instabilities differ from common nonlinear instable regimes observed in nonlinear optics, since here the feedback is provided by the scattering. According to Skipetrov and Maynard \cite{SkipetrovPRL00,SkipetrovPRE01}, the instability threshold $p ={\langle n_2I\rangle}^2(L/\ell)^3 > 1$ is reached for any value of the nonlinear correction to the refractive index $n_2I$, as long as the mean free path $\ell$ is small enough. To date, experimental observation of speckle instability has never been reported.

In this letter, we report the first experimental evidence of speckle instability. A liquid crystal (LC) cell serves simultaneously as the scattering and Kerr nonlinear medium. A two dimensional (2D) computer-generated image of random scatterers is projected on a photoconductive wall of the cell, generating a controlled 2D random distribution of the refractive index within the LC layer. A probe laser beam is transversally scattered while inducing a nonlinear (NL) Kerr-reorientation of the LC molecules. Speckle pattern spontaneous oscillations are observed. Instability threshold is identified and found to be dependent on scattering mean free path, as predicted by theory. A frequency cascade is observed at higher probe intensity. Just above threshold, the oscillation frequency is found to be of the same order of the inverse of the NL relaxation time,  which is interpreted in terms of NL energy transfer and natural gain selection. 
Light scattering in random and nonlinear media has attracted considerable attention recently with, for instance, the exploration of the impact of a nonlinearity on Anderson Localization \cite{Segev07,Silberberg08} or the role of disorder on solitons \cite{Solitons}. Our experimental study is among the very first \cite{Cao03} to explore the dynamic interplay between nonlinearity and multiple scattering.


\begin{figure}[h!]
\centering
\includegraphics[width=70mm]{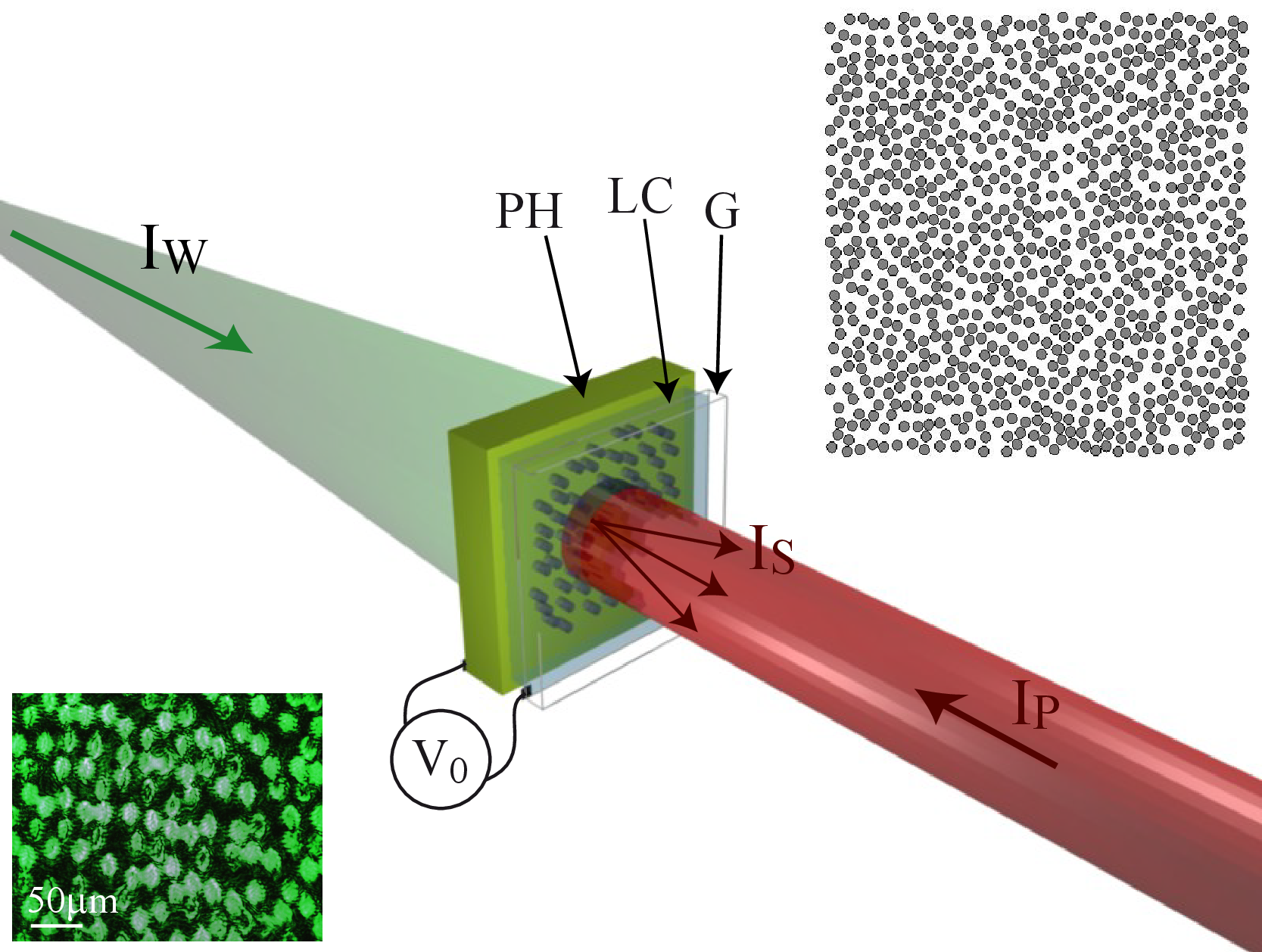}
\caption{(color online). Experimental configuration. The disorder configuration is imprinted within the LC (bottom-left inset) by illuminating ($I_W$) the photoconductive wall ($PH$) of the LC light-valve with a random distribution of scatterers (top-right inset), whereas the nonlinearity is provided by the molecular reorientation of the liquid crystal under the action of the probe probe ($I_P$). (G=glass cover).}
\label{fig1}
\end{figure}

The experimental setup is schematically represented in Fig.\ref{fig1}. The 20mm$\times$30mm liquid crystal light-valve (LCLV) is composed of a $L = 55 \mu$m-thick LC film in the nematic phase and a 1 mm-thick $Bi_{12}SiO_{20}$ (BSO) photoconductive substrate, sandwiched between two transparent indium-tin-oxide electrodes \cite{Stefania06}.
A random pattern (here a random collection of dark disks in a white background, as shown in top-right inset of Fig.~\ref{fig1}) is computer-generated and transmitted to a 36.9mm$\times$27.6mm spatial-light-modulator (SLM), spatial resolution 1024$\times$768 pixels, which serves as a 2D random mask. The local voltage applied across the LC film decreases where light impinges on the BSO crystal, inducing local LC reorientation. A given light intensity distribution is therefore converted into a corresponding distribution of molecular orientation. As a result of LC birefringence, a 2D refractive index distribution invariant along the longitudinal direction, $n(x,y)$, is formed within the LC film, which reproduces the initial random pattern. Because the BSO crystal is preferentially sensitive to the blue-green region of the spectrum, a CW diode-pumped solid state laser at $532$ nm is used as the writing beam with an intensity of a few mW/cm$^2$ (2mW/cm$^2$). The refractive index distribution is monitored by a 8-bit CCD-camera and is shown in bottom-left inset of Fig.\ref{fig1}. The geometrical characteristics of the LC scattering medium are fully controlled by adjusting the filling fraction, $\phi$, the diameter $D$ and positions of the cylindrical scatterers. By shrinking the image on the computer screen, the actual scatterer diameter can be varied from $D=100 \mu$m to 1 $\mu$m, leaving $\phi$ constant. The refractive index of the scatterers, $n_s$, varies between $n_o=1.52$ and $n_e=1.75$, according to the 8-bit coded gray level of the SLM mask, where $n_o$ and $n_e$ are, respectively, the ordinary and the extraordinary index of the LC. In the following, we choose $\phi=$40\%, $D$ = 50 or 20 $\mu$m and $n_s=n_e=1.75$ while for the LC matrix $n=n_o$.

\begin{figure}[h!]
\centering
\includegraphics[width=70mm]{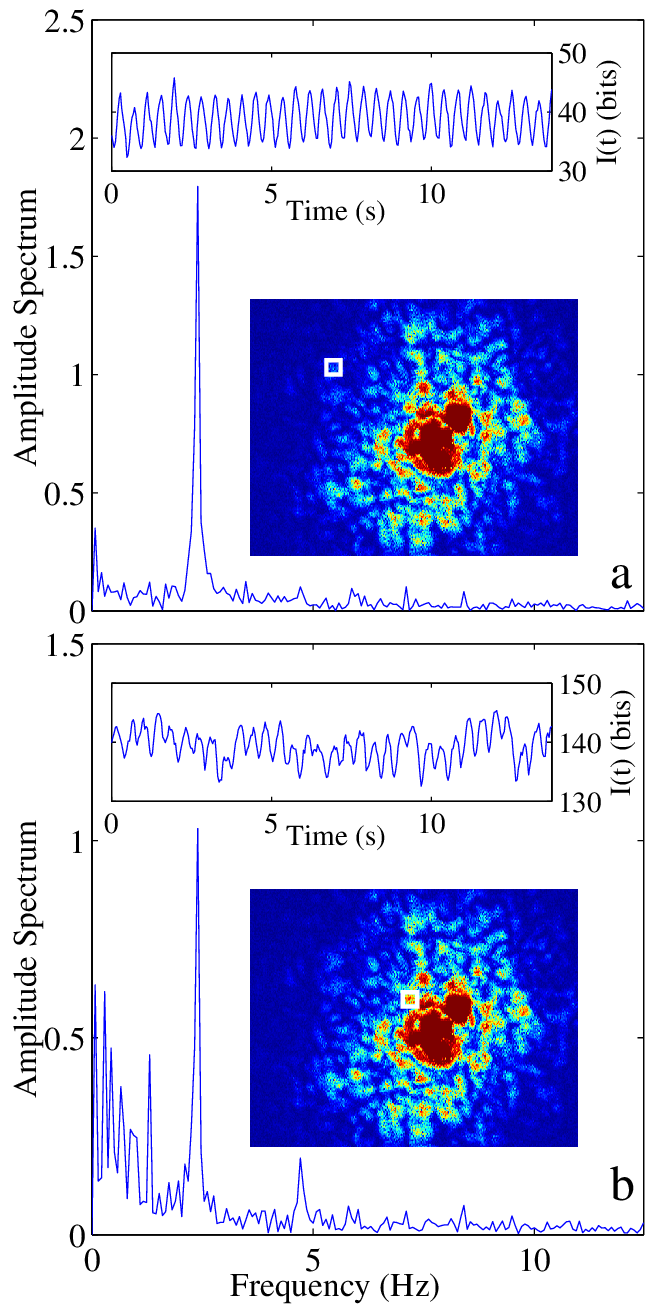}
\caption{(color online). Temporal modulation of speckle intensity and corresponding power spectrum within an area encompassing a single speckle spot (delineated by the white square shown in the speckle pattern CCD image). Time-averaged intensity in this area is (a) I=40 bits and (b) I=140 bits.}
\label{fig2}
\end{figure}

A linearly polarized HeNe laser (632.8nm wavelength, 5mW power, beam-waist $w_0$=380 $\mu$m, intensity 1.1 W/cm$^2$) serves as the probe beam. Its polarization is parallel to the LC nematic director. The probe beam is transversally scattered as it propagates along the LC layer and is back-reflected at the LC/BSO interface with a total reflectivity of 22\%. The back-reflected near-field speckle pattern is imaged onto a second CCD camera and is shown in Fig.~\ref{fig2}. We checked that polarization is conserved. This confirms that propagation remains in the paraxial approximation limit and allows to estimate the mean free path in the transverse direction. Indeed in this limit, the wave equation reduces to a Schr\"{o}dinger equation where time is replaced by longitudinal coordinate, $z$, to describe a particle moving in a 2D potential $-V(x,y)=-k^2[n^2(x,y)-n^2_0]$:
\begin{equation}
2ikn_0\frac{\partial A}{\partial z}=(\nabla^2_{xy}+V(x,y))\phi
\end{equation}
where $A(x,y,z)\exp(-ikn_0z)$ is the wave field and $n_0$ is the effective refractive index defined by $n_0=S^{-1}\int_Sn^2(x,y)dxdy$ \cite{DeRaedt89}. When considering transverse scattering, the relevant wavewavector is no longer $k=\omega/c$ but rather the transverse wave-vector $k_\bot=\omega/v=2/w_0$, with $v/c=k/k_\bot\approx2$ in our case. Because this ratio can be larger than 1, the product $k_\bot\ell\ll k\ell$, making transverse Anderson localization of light possible in finite-sized random systems even for small index contrast \cite{DeRaedt89,Segev07,Silberberg08}. In our system, a crude estimate of the mean free path from Mie theory (infinite cylinders with index $n_e$ in a matrix $n_o$) and single scattering approximation yields $\ell\approx100\mu$m (resp. $\ell\approx40\mu$m) at 632.8nm for $D=50\mu$m (resp. $D=20\mu$m). The localization length in 2D is given by $\xi\approx\ell\exp(\pi k_\bot\ell/2)$, which yields $\xi=228,6\mu$m (resp. $\xi=55.7\mu$m). As it propagates freely along the longitudinal direction, the beam will extend transversally, experiencing successively a ballistic regime ($t<\tau_0=\ell/v$), a subdiffusive multiple-scattering regime ($\tau_0<t<T_{loc}$) \cite{Sebbah93} and a crossover to localization ($t>T_{loc}$). To identify the regime reached within the round-trip ballistic time $t_b=2L/c$ spent in the LC cell, we estimate $t_b/\tau_0=(d/\ell)(k/k_\bot))\approx2.2$ (resp. 5.5), and $t_b/T_{loc}=2(t_b/\tau_0)exp(-\pi k_\bot\ell)\approx0.8\%$ (resp. 5.7). Although based on a rough estimate of the mean frfee path, these calculations demonstrate that the multiple scattering regime is well established, even though light may not be yet fully transversely localized.

\begin{figure}[h!]
\includegraphics[width=70mm]{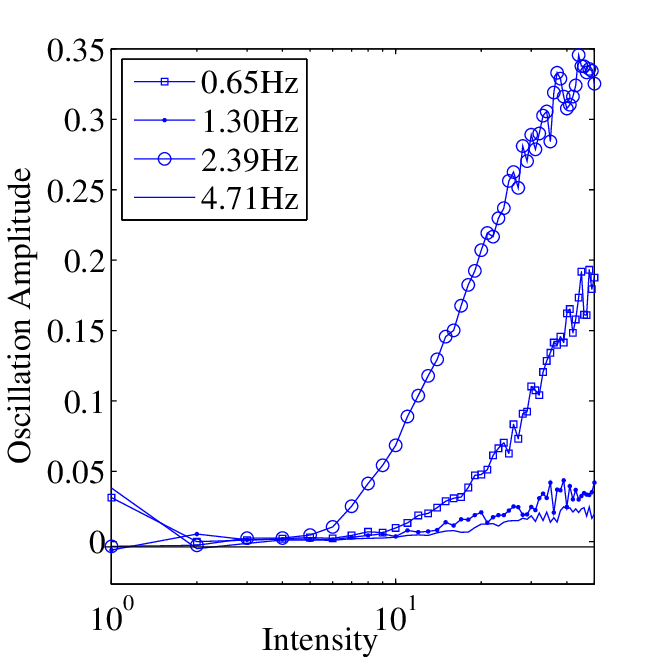}
\caption{(color online). Amplitude of oscillations vs average intensity at the different frequencies seen in Fig~\ref{fig2}b, obtained by averaging spectra measured at positions of the speckle pattern with same time-averaged intensity.}
\label{fig3}
\end{figure}

During its round-trip propagation in the cell, the extraordinary-polarized probe beam experiences nonlinear scattering as its electric field changes molecule orientation. Surprisingly, the speckle pattern is not static but we observe with a naked-eye a temporal modulation of each speckle spot (as shown in the movie \cite{movie}). These oscillations are not observed for ordinary-polarized light and therefore cannot be attributed to thermal fluctuations. They are found to average out when spatial integration is performed, which shows that no phase relation exists from spot to spot. It is important to point out that our observations are fundamentally different from standard self-phase modulation (SPM) described in the literature for non-disordered media \cite{SPM}. The uncorrelated oscillations for different speckle spots confirm the non-deterministic nature of the speckle instability, in contrast for instance with the phase-coherent rings observed in SPM.

We analyze the oscillating speckle behavior by monitoring over time the modulated intensity, spatially averaged over a small area including a single speckle spot. The temporal oscillations and the corresponding amplitude spectrum are shown in Fig.~\ref{fig2} for two speckle spots with different time averaged intensity. Low intensity speckle spot shows a single peak at $f_0$=2.39 Hz (Fig.~\ref{fig2}a). This has to be compared to the free relaxation time of the reorientational nonlinearity of the LC, $\tau_{NL}=\gamma L^2/\pi^2K$, where $\gamma$ is the LC rotational viscosity and $K$ the elastic constant. For typical values of $\gamma=0.02$ Pa.s and $K=15$ pN \cite{Zeldovich,Khoo-physrep}, we find with good agreement $f_0\approx1/\tau_{NL}$. The same time constant, approximately $400$ ms, has been independently found by measuring directly the response time of the optical reorientational effect. The speckle instability is therefore solely driven by the slow response time of the reorientational Kerr nonlinearity. This is another reason why it cannot be related to thermal instabilities since thermal fluctuations are expected to be faster than molecular orientation of the LC  (of the order of 100 $\mu$s \cite{Khoo-physrep}). Besides, given that the intensity used is relatively small and given the small absorption coefficient of pure LC \cite{Khoo-physrep}, we can safely neglect thermal heating due to laser illumination.

\begin{figure}[h!]
\includegraphics[width=70mm]{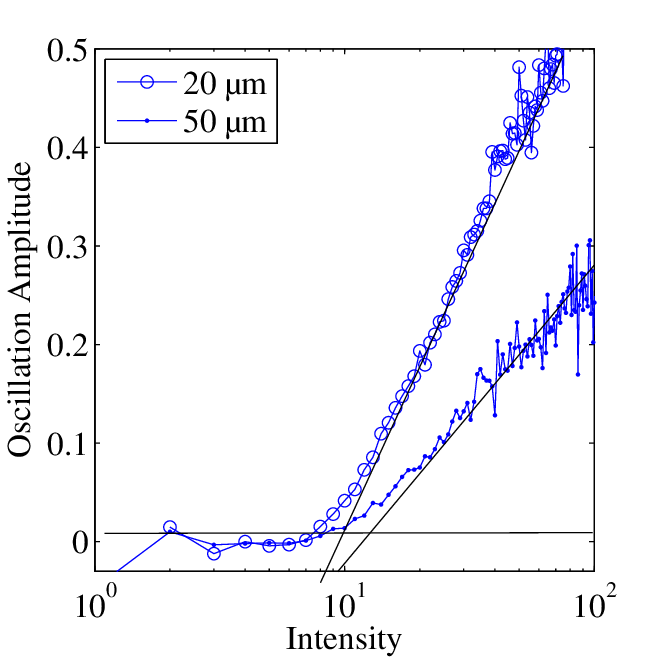}
\caption{Amplitude of oscillations at $f_0$=2.39 Hz vs. average intensity for two different scatterer diameters $D$ = 50 $\mu$m and 20 $\mu$m, corresponding respectively to scattering media with mean free path $\ell\approx100\mu$m and $\ell\approx40\mu$m. The threshold, which is materialized by the crossing of the lines, decreases with decreasing $\ell$.}
\label{fig4}
\end{figure}

Higher intensity speckle spots show several spectral peaks at $f_0/2$, $f_0/4$ and also at $2f_0$ (Fig.~\ref{fig2}b). Increasing the probe beam intensity results in the same spectral cascade. Remarkably, the intensity distribution of the speckle pattern offers a direct exploration of the intensity dependence of the instability at fixed input beam intensity. This is further analyzed by averaging spectra measured at positions with same time-averaged intensity. Four peaks with different threshold are identified and their amplitudes relative to the spectral background are plotted versus speckle spot intensity in Fig.~\ref{fig3}. Different slopes associated with different threshold are observed for each oscillation frequency. Next we investigate the instability threshold dependence with disorder. According to the theoretical predictions \cite{SkipetrovPRL00,SkipetrovPRE01}, the instability threshold $p ={\langle n_2I\rangle}^2(L/\ell)^3 > 1$ is expected to decrease with decreasing mean free path. We compare the intensity dependence of the peak amplitude at the main oscillation frequency $f_0$ for scatterers with diameters $D$ = 50 $\mu$m and 20 $\mu$m, corresponding respectively to $\ell\approx100\mu$m and $\ell\approx40\mu$m. Changing the scatterer dimensions does not affect the oscillation frequency $f_0$. However the instability threshold is clearly seen in Fig.~\ref{fig4} to decrease with increasing scattering (decreasing $\ell$). This clearly demonstrates the essential role of disorder in speckle instability and the threshold dependence on the mean free path, as predicted by theory. A more systematic study of the threshold dependence with disorder and a comparison with theoretical predictions are under investigation.

We interpret these experimental observations in terms of parametric amplification. In a parametric process, a pump beam at $f_p$ and a probe beam at $f_m$ interfere to create a grating within the nonlinear material. The pump beam is itself scattered by this grating and nonlinearity-driven energy transfer is achieved from the pump to the probe when $\Delta f=f_p-f_m\neq0$. This transfer of energy via the Kerr nonlinearity has been interpreted as an amplification process, where the gain reads as $G(\Delta f)=2\Delta f\;\tau_{NL}[1+(\Delta f\;\tau_{NL})^2]^{-1}$ \cite{Silberberg84}. The corresponding gain curve is peaked at $\Delta f=1/\tau_{NL}$. Starting from noise, energy from the pump at $f_p$ will be preferentially transferred into photons at $f_m=f_p-1/\tau_{NL}$ resulting in the beating of the two beams. In the same way, the spontaneous speckle oscillations observed in our experiment result from the beating between the incident HeNe beam and the nonlinearly-generated beam at the maximum of the gain curve, giving rise to a temporal modulation at $f_0\approx1 /\tau_{NL}$.

In conclusion, we have observed speckle instabilities in transverse scattering of light in random LC films with reorientational Kerr effect. The instability is solely attributed to the combination of scattering and Kerr nonlinearity, with uncorrelated oscillations for different speckle spots. It depends on polarization, excluding any thermal effect. The intensity distribution of the speckle allows a direct probing of the intensity-dependence of the instability spectrum. A threshold is found which depends on the degree of disorder ($\ell$). This confirms earlier theoretical predictions \cite{SkipetrovPRL00}. The oscillation frequency just above threshold is equal to the inverse of the NL relaxation time and is interpreted in terms of the beating between the incident beam and a generated light emitted at a frequency equal to the maximum of the gain curve associated with the Kerr effect. Two regimes of speckle instability have been predicted theoretically \cite{SkipetrovJOSA04}: a slow NL regime where $\tau_{NL}>\tau_D$ driven by the relaxation time of the non linearity, as in the experiment presented here, and a fast NL regime when $\tau_{NL}<\tau_D$. This second regime was out of reach in the present experiment since the reorientational Kerr NL is extremely slow. However, we conjecture that a frequency of oscillation different from $1/\tau_{NL}$ should be observed, which will strongly depend this time on the characteristics of the scattering medium. As in a random laser \cite{Vanneste01,AndreasenAOP10}, the gain curve will select an eigenmode of the passive system, which in turn will beat with the pump. Different solutions can be suggested to reduce the relaxation time of LC NL reorientation, using for instance ferroelectric liquid crystal ($\tau<1$ms). This would allow the exploration of the fast NL regime ($\tau_{NL}<\tau_D$), where the instability oscillation should depend on the scattering characteristics of the material.

This work was supported by the Federation D\"{o}blin FR2008 and the French National Research Agency under grant ANR-08-BLAN-0302-01.

{}

\end{document}